\documentstyle[psfig,raulbib,times]{mn}

\begin{document}

\title[Archetypal analysis of galaxy spectra]
{Archetypal analysis of galaxy spectra}
\author[Chan, Mitchell \& Cram]{
B.H.P. Chan, D.A. Mitchell and L.E. Cram\\
Astrophysics Department, School of Physics, A28, University of Sydney, NSW 2006, Australia\\
\textit{email:} bchan; mitch; lc @physics.usyd.edu.au\\
}
\maketitle

\begin{abstract}
Archetypal analysis represents each individual member of a set of data
vectors as a mixture (a constrained linear combination) of the 
\textit{pure types} or \textit{archetypes} of the data set. The 
archetypes are themselves required to be mixtures of the data vectors. 
Archetypal analysis may be particularly useful in analysing data sets 
comprising galaxy spectra, since each spectrum is, presumably, a 
superposition of the emission from the various stellar populations, 
nebular emissions and nuclear activity making up that galaxy, and each 
of these emission sources corresponds to a potential archetype of the 
entire data set. We demonstrate archetypal analysis using sets of 
composite synthetic galaxy spectra, showing that the method promises 
to be an effective and efficient way to classify spectra. We show that
archetypal analysis is robust in the presence of various types of noise.
\end{abstract}


\section{Introduction}
Archetypal analysis is a statistical data representation technique
developed by \scite{Cutler94} to characterise multivariate data sets 
of the form $\{ {\bf x}_i, i = 1, 2, .... n\}$, where each 
$\{{\bf x}_i\}$ is an $m$-vector with $m$ variables, i.e. ${\bf x}_i 
= (x_{i1}, x_{i2}, .... x_{im})$. The algorithm represents every member 
of the set $\{{\bf x}_i\}$ as a mixture (a constrained linear combination) 
of basis vectors that are \textit{pure types} or \textit{archetypes} of 
the data set. The archetypes are themselves mixtures of the data vectors. 
Our exploratory study of archetypal analysis of galaxy spectra is 
justified by noting that the spectra of many galaxies appear to be a 
superposition of emissions from various \textit{populations} or 
\textit{mechanisms}. Identification of different archetypes with the 
spectral signature of these different populations or mechanisms promises 
to provide a natural representation of galaxy spectra.

Data representation techniques previously applied to galaxy spectra
include principal component analysis (PCA) (Ronen et al. 1999, and
references therein), multiple optimised parameter estimation and data
compression (MOPED) \cite{HJL00,RJH01} and the information
bottleneck method (IB) \cite{Slonim01}. An objective of these
representation techniques has been to identify the physical 
relationships between different emission signatures, and to attempt to 
uncover parameterizations describing these physical relationships.
Additionally, data representation techniques frequently produce a more
compact representation of the data set and thereby compress the
computation and data storage load. A critical review of both aspects
of PCA, MOPED and IB is given by \scite{Lahav01}.

The eigenvectors determined by principal components analysis and its
relatives generally do not resemble any member of the data and may be
difficult to interpret. To overcome this, Cutler \& Breiman required
that the basis vectors in archetypal analysis (the archetypes) be
\textit{mixtures} of the actual data points. In particular, archetypes
are \textit{extreme members} of the data set chosen in the following
sense. Consider the $m$-dimensional polytope (a region of $m$-dimensional 
space enclosed by a set of hyperplanes) corresponding to the convex hull 
(the hypersurface defined by the minimum number of extreme data points)
of the $n$ data points (hereinafter called the \textit{data convex hull}). 
The vertices of this polytope are by definition data points, and as proven 
by \scite{Cutler94} any point on the polytope can be represented as a 
mixture of the data points. Archetypes are 
chosen from the data mixtures lying on the data convex hull. 
Since the typical number of archetypes used is smaller (generally much 
smaller) that the number of vertices of the data convex hull, the
$m$-dimensional polytope corresponding to the convex hull of the
archetypes (hereinafter the \textit{archetype convex hull}) encompasses a
smaller hypervolume than the data convex hull. Data points lying
inside the archetype convex hull are exact mixtures of the archetypes, 
while data points lying outside the archetype convex hull are only 
approximated. The difference reflects a loss of information arising from 
the use of a reduced representation, and is similar to the loss arising 
from truncating the number of eigenvectors used in PCA.

In Section 2 of the paper we present a more extensive description of
archetypal analysis. In Section 3 we demonstrate the application of
archetypal analysis to extract the star formation history from model
galaxy spectra and also the effects on archetypal analysis in the
presence of various types of noise. Finally, a discussion of future
work is presented in Section 4.


\section{Archetypal analysis}
Here we contextualise the mathematical basis of archetypal analysis
using galaxy spectra, based on the development in
\scite{Cutler94}. From a starting point of principal components
analysis, we first present the mathematical methodology of archetypal
analysis, and then use a schematic example to illustrate the
differences between PCA and archetypal analysis.

Suppose that we are given a set of $n$ galaxy spectra $\{{\bf x}_i, 
i=1, ...,n\}$, each consisting of a measurement (e.g. of flux density)
in $m$ wavelength bins, i.e. ${\bf x}_i = (x_{i1},...,x_{im})$.  For
any given set of $p$ $m$-vectors $\{{\bf z}_k, k=1,...,p\}$
(hereinafter the basis vectors), the linear combination
$\sum_{k=1}^{p} \alpha_{ik}{\bf z}_k$ that best approximates any given
galaxy spectrum in the data set ${\bf x}_i$ may be defined through the
coefficients ${{\alpha_k}}$ that minimize the error
\[
\parallel {\bf x}_i - \sum_{k=1}^{p} \alpha_{ik}{\bf
z}_k \parallel ^2.
\]
The ${{\alpha_k}}$ minimize the squared distance between the data 
point ${\bf x}_i$ and the represented point $\sum_{k=1}^{p}
\alpha_{ik}{\bf z}_k$ in $m$-dimensional hyperspace.
The optimal set of basis vectors $\{{\bf z}_k\}$ to represent the
entire data set may then be determined as the minimizer of the sum of
all squared distances over the entire data set,
\begin{equation}
\sum_{i=1}^{n} \parallel {\bf x}_i - \sum_{k=1}^{p}
\alpha_{ik}{\bf z}_k \parallel ^2.
\label{equ:MIN}
\end{equation}
Suppose that the set of basis vectors $\{{\bf z}_k\}$ is chosen to be 
an orthonormal set of axes. Then expression~\ref{equ:MIN} minimizes the 
sum of the distances from the data points to the axes. This is equivalent 
to maximizing the sum of the projections onto the axes (cf. fig. 2.3, 
Murtagh \& Heck 1987) given by, 
\begin{equation}
\sum_{k=1}^{p} {{\bf z}^t}_k {\bf \sf S} {\bf z}_k,
\label{equ:MAX}
\end{equation}
where {\bf \sf S} is the variance-covariance matrix of the data set,
\[
{\bf \sf S}_{ij} = \frac{1}{n} \sum_{l=1}^{n}(x_{li} - \bar{x}_i)(x_{lj} - \bar{x}_j).
\]
The minimizers of expression~\ref{equ:MAX} are the eigenvectors of 
{\bf \sf S} corresponding to the $p$ largest eigenvalues. When the data 
are centred on the mean ($\bar{x}_j = \frac{1}{n} \sum_{i=1}^{n}x_{ij}$),
this corresponds to the principal component decomposition of principal 
component analysis (PCA).

A basis vector derived by such principal component decomposition does
not necessarily correspond to, or even resemble, any member or
combination of members of the original data set. Furthermore, it is
not required that each data point is approximated as a mixture of the
basis vectors, only that the representation globally minimizes the
residual sum of squares. These features may make the physical
interpretation of the principal components difficult.  For instance,
in applying PCA to a set of galaxy spectra, the principal components
(eigenspectra) may not resemble the spectrum of any observed galaxy,
and may contain points that are negative, or which vary rapidly and
unphysically with wavelength. Furthermore, the representations of
individual data points may contain negative or wildly varying
weightings of the basis vectors.

As noted by \scite{ML01}, these problems can be overcome to some
extent. These authors applied PCA to a set of galaxy spectra obtained
in the 2dF galaxy redshift survey \cite{Colless01} and then formulated
physically significant parameterizations from linear combinations of
the first two eigenspectra. As they stated: ``\textit{In effect what 
we are doing when we utilise these linear combinations is rotating the 
axes defined by the PCA to make the interpretation of the components 
more straightforward}''. The axes defined by PCA potentially require 
rotation because PCA itself determines orthogonal (linearly
independent) eigenspectra, while the physical effects that dominate
changes in the appearance of galaxy spectra are not necessarily
independent in this way.

In contrast to PCA, archetypal analysis was designed with two distinct
features that address the question of physical interpretation more
directly: (1) the basis vectors are themselves required to be members
or mixtures of members of the input data set, making an interpretation
of the origin of the basis vectors more straightforward, and (2) the
basis vectors are extreme data points lying on the data convex hull,
allowing each member of the data set to be represented either exactly
or approximately as a mixture of the basis vectors. These properties 
are achieved by introducing two additional sets of constraints to the 
general formalism of PCA. Begin by selecting a set of basis vectors 
$\{{\bf z}_k, k=1,...,p\}$ that are ``linear combinations'' of the data 
values,
\[
{\bf z}_k = \sum_{j=1}^{n} \beta_{kj}{\bf x}_j; \hspace{2em}k = 1,...,p.
\]
with \(\beta_{kj} \geq 0\) so the basis vectors ``resemble'' the data, 
and \(\sum_j \beta_{kj} = 1\) so that they are a ``mixture'' of the data. 
Then determine coefficients $\{\alpha_{ik}\}$ that allow the set of data 
points to be ``well represented'' by the basis vectors via minimizing
\[
\parallel {\bf x}_i - \sum_{k=1}^{p} \alpha_{ik}{\bf
z}_k \parallel ^2.
\]
Finally, use the constraints $\alpha_{ik} \geq 0$ so that each data
point is a ``physically meaningful'' combination of basis vectors, and
$\sum_k \alpha_{ik} = 1$ so that data points are ``mixtures'' of
basis vectors. The set of basis vectors $\{{\bf z}_k\}$ that optimally
describes the data set are then minimizers of the \textit{residual sum of
squares},
\[
\sum_{i=1}^{n} \parallel {\bf x}_i - \sum_{k=1}^{p}
\alpha_{ik}{\bf z}_k \parallel ^2,
\]
or
\begin{equation}
\sum_{i=1}^{n} \parallel {\bf x}_i - \sum_{k=1}^{p}
\alpha_{ik} \sum_{j=1}^{n} \beta_{kj}{\bf x}_j \parallel ^2.
\label{equ:AAeq}
\end{equation}

The basis vectors determined by these sets of constraints are called
{\it archetypes}.  An archetypal representation involves the
determination of the two sets of coefficients ${\alpha_{ik}}$ and
${\beta_{kj}}$. Cutler \& Breiman prove that archetypes lie on the
data convex hull, and show how the coefficients can be determined
iteratively using an alternating constrained least-square
algorithm. As they note, unlike PCA, archetypal analysis does not
nest, nor are archetypes orthonormal, so that existing archetypes
change to capture in progressively better ways (subject to the advent
of noise) the shape of the data set as a larger set of archetypes is
determined.

\begin{figure}
\centering{\psfig{file=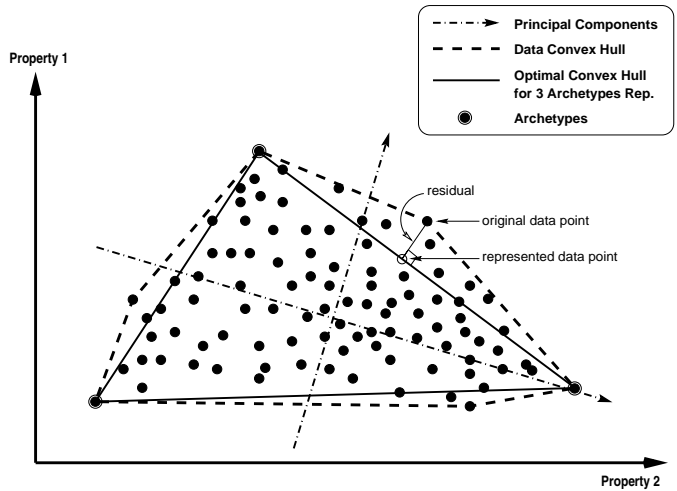,width=.5\textwidth,angle=-90}}
\caption{Diagrammatic comparison of archetypal analysis and PCA. The
two dot-dashed axes illustrates the principal components of PCA. The 
dashed hexagon is the data convex hull, and the solid triangle shows 
the archetype convex hull for the case of three archetypes. Points 
lying within the archetype convex hull are represented exactly, while 
points lying outside are approximated by the nearest point on the 
archetype convex hull, as illustrated.}
\label{fig:AAvsPCA}
\end{figure}

\subsection{An exemplar}
It is instructive to visualise the difference between archetypal
analysis and PCA using an artificial two-dimensional (i.e. $m=2$) 
data set. Such a data set is exhibited as the points in
Fig. 1.  Using PCA, the data are represented by
their projections (coordinates) referenced to principal components
pointing in the two orthogonal directions of maximum data variance,
illustrated as the dot-dashed axes. Because there are only two
independent variables, two principal components will provide an exact
reconstruction of every data point.

Archetypal analysis is a representation using basis vectors lying on
the data convex hull. In Fig. 1 the dashed hexagon
traces the data convex hull, like a rubber band stretched around the
data set. The archetypal algorithm locates archetypes on this data
convex hull that form (for the given number of archetypes) an optimal
model of the convex hull. For example, if we represent the artificial
data set with three archetypes, the archetypal algorithm determines 3
points on the data convex hull that outline a triangle, which is the
archetype convex hull. Every data point within the triangle is exactly
described (zero residual) by mixtures of the archetypes. Data points
lying outside the triangle are represented as the nearest point on the
triangle, as illustrated in Fig. 1. The distance between the original 
(external) data point and the represented data point is the 
\textit{residual}. The sum of these residuals is minimized 
(expression~\ref{equ:AAeq}) in the search for the optimal set of 
archetypes. For this particular data set, the data convex hull is a 
hexagon and six archetypes would provide an exact reconstruction of 
every data point.

There are three generic issues associated with archetypal analysis that 
are addressed in the following sections:
\begin{enumerate}
\item The quality of an archetypal representation of multi-dimensional
data depends on the number of archetypes adopted. We demonstrate the
outcome of using different numbers of archetypes.
\item Archetypes are located on the data convex hull and archetypal
analysis is potentially sensitive to outliers. We investigate the
effects of noise-generated outliers on archetypal analysis.
\item The input data set sometimes needs to be standardised before
archetypal analysis. We discuss our experience in this respect.
\end{enumerate}

\section{Application to synthetic spectra}
We apply archetypal analysis to study evolving star formation in model
galaxies. We construct a set of spectra of model galaxies by assuming
that each galaxy has experienced episodes of star formation with
quasi-random amplitudes at past epochs. Motivated by the analysis of
Reichardt et al. (2001), we model the galaxies as superpositions of stellar
populations formed instantaneously, and now observed at ages of 0.007,
0.045, 0.3, 2 and 14 Gyr. The {\it fiducial spectra} at
these ages are derived using the {\sc P\'{e}gase-II} code of
\scite{Fioc99}, adopting a \scite{Salpeter55} initial mass function in
the range [0.1,120] ${\rm M_\odot}$ and the stellar library of
\scite{LCB9798}. Spectra were synthesised over the optical wavelength
range 365 -- 711 nm, with 2 nm spectral bins (174 bins). Since the
shapes of the nebular emission lines are not predicted by {\sc
P\'{e}gase-II} we simulate them using Gaussians with a FWHM $\sim$ 5 nm. 
This treatment is not required for archetypal analysis, but allows
convenient visual presentation.

\begin{figure}
\centering{\psfig{file=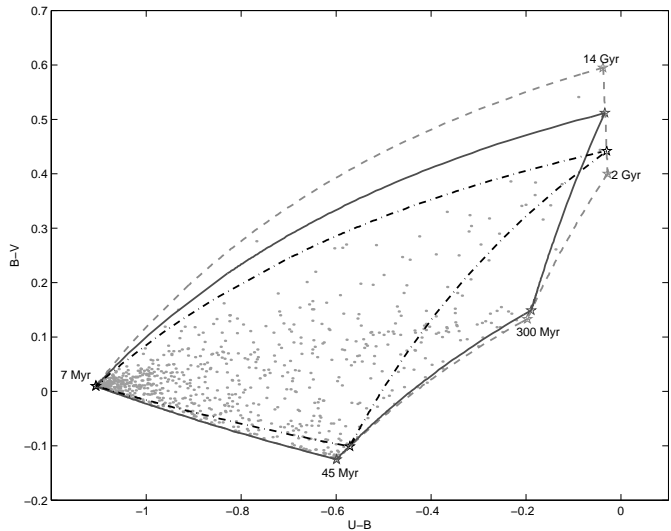,width=.5\textwidth,angle=90}}
\caption{The projection of the data convex hull and the data
themselves of a set of 900 noise free, composite spectra onto the
two-colour $(B-V,U-B)$ plane.  The dashed line outlines the data convex
hull, the solid line shows the archetype convex hull for four
archetypes, and the dot-dashed line shows the archetype convex hull for
three archetypes.  }
\label{fig:ATconvexhull}
\end{figure}

To generate a model composite spectrum, the five fiducial spectra are
first normalised to the same integrated flux. A composite spectrum is 
then generated via the superposition of fiducial spectra using a set of 
five quasi-random weighting factors. To satisfy the constraint that the 
sum of the five weighting factors be unity whist having quasi-uniform 
distribution, we generate the weighting factors according to the relationship 
$w_i = {\rm U}[0,1] (1-\sum_{n=1}^{i-1}w_n)$, where ${\rm U}[0,1]$ 
denotes an uniformly distributed random number in [0,1]. Up to 5 random 
numbers are generated this way, or they are set to zero when 
$\sum_{n=1}^{i-1}w_n \geq 1$. These pseudo-random values are then randomly 
assigned as the five weighting factors. Since one of the five random values 
is never scaled, it retains an uniform distribution within [0,1] while 
the sum is constrained to be unity.

The sample of model galaxies is generated by repeating the process with 
different random numbers for each galaxy. In addition to determining the 
spectrum for each model galaxy, we also calculate the standard photometric 
colours. The distribution of the model galaxies in the two-colour $(B-V,U-B)$
plane is shown in Fig. 2. The fiducial spectra themselves are displayed in 
the first 5 panels across the top row of Fig. 3. The sixth panel exhibits 
an example of a composite spectrum, composed from ratios of 0.14, 0.66, 
0.00, 0.13, and 0.08 of the fiducial spectra.

\begin{figure*}
\centering{\psfig{file=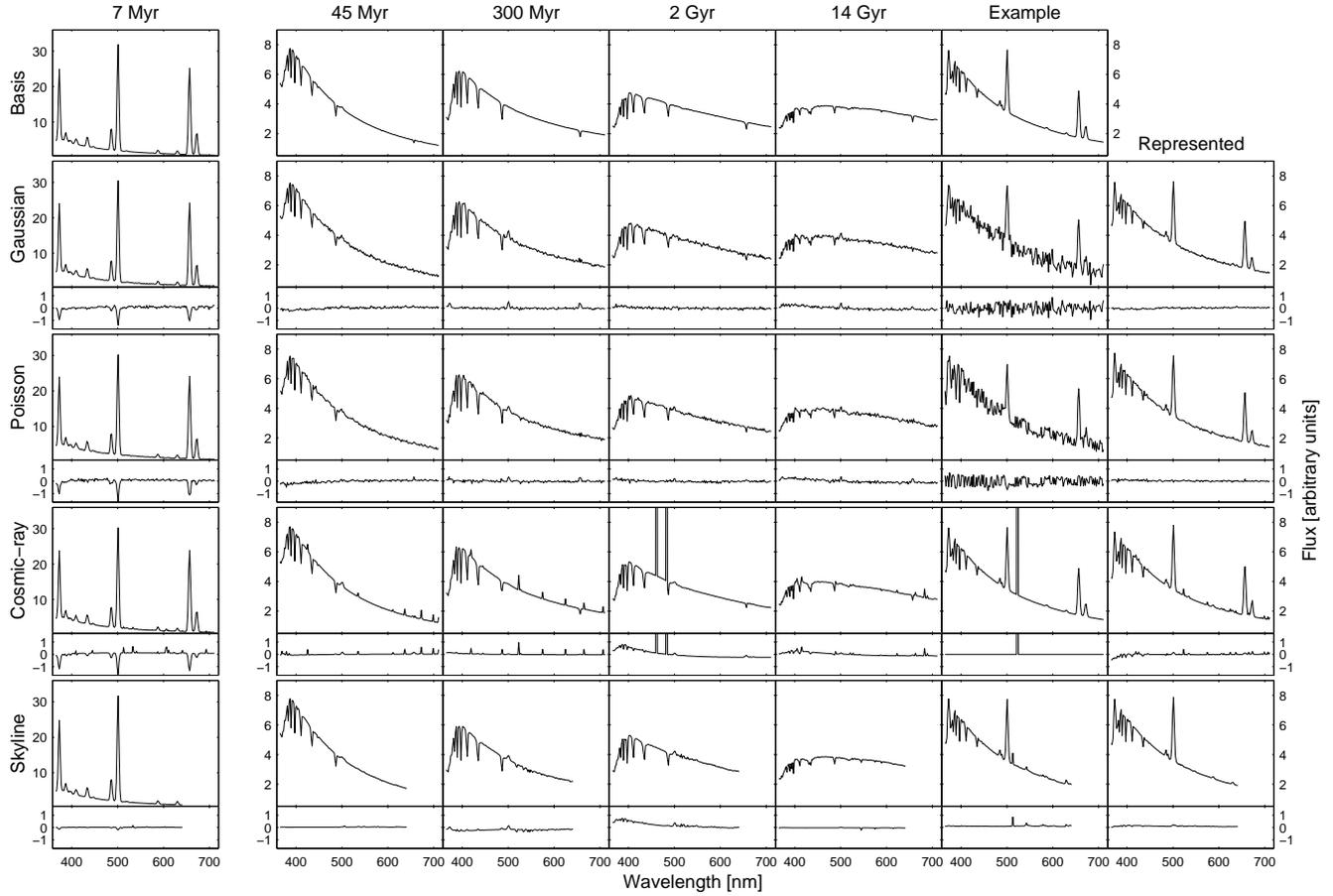,width=\textwidth,angle=90}}
\caption{Top row: Fiducial spectra (columns 1-5), with a sample
composite spectrum (column 6). Rows 2-4: Illustrating respectively the
effects of model Gaussian (S/N=10), Poisson (S/N=10) and cosmic ray
noise (5\% random spikes), and night sky residual (with 5\% amplitude)
when five archetypes are used. In each row, columns 1-5 are the
recovered archetypes, column 6 is the composite spectrum with its
error, and column 7 is the represented composite spectrum. The small
panels below rows 2-4 illustrate the differences between the spectrum
immediately above, and the noise free counterpart in row 1.}
\label{fig:ATs}
\end{figure*}

\begin{figure*}
\centering{ \mbox{
\psfig{figure=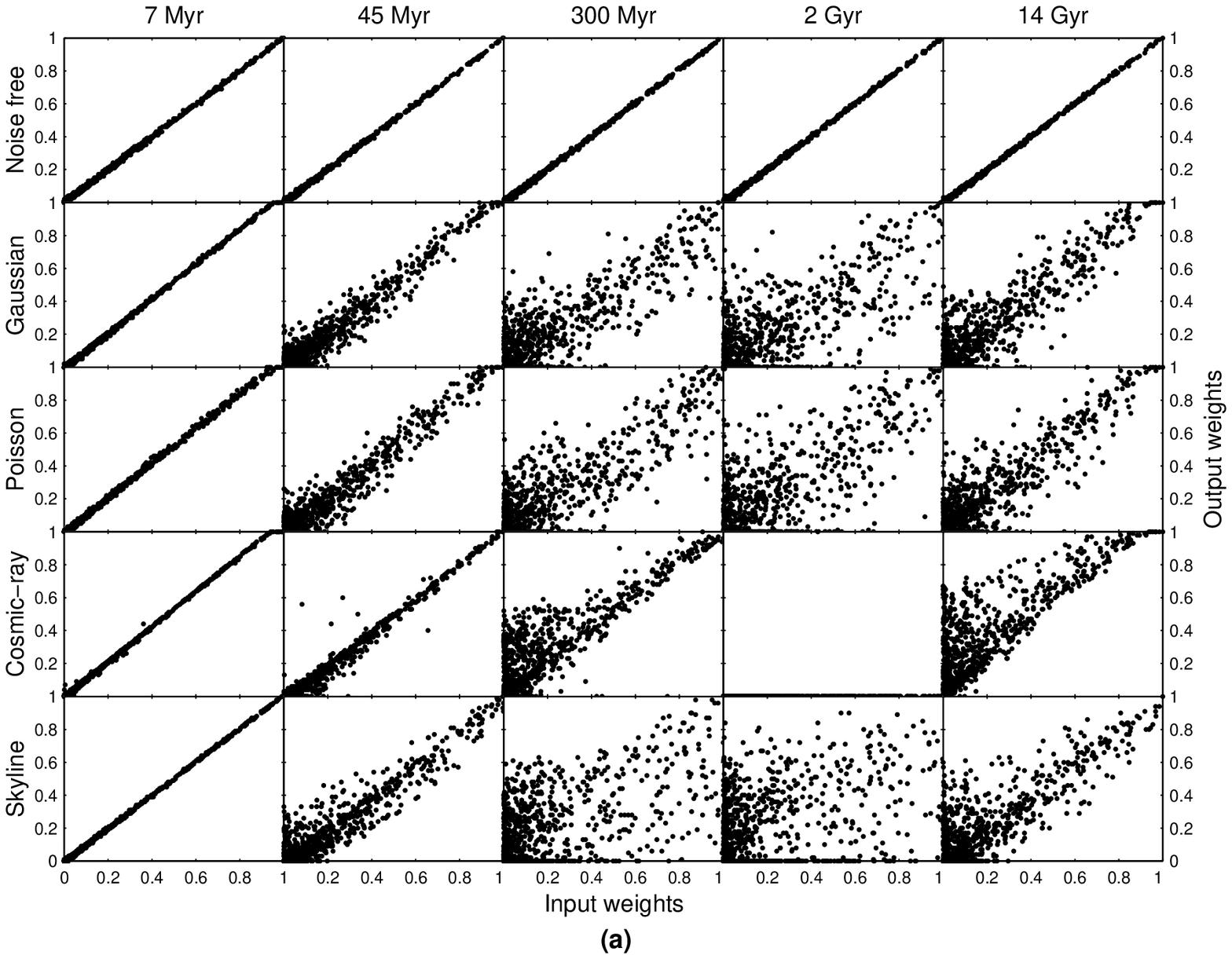,width=.5\textwidth}\quad
\psfig{figure=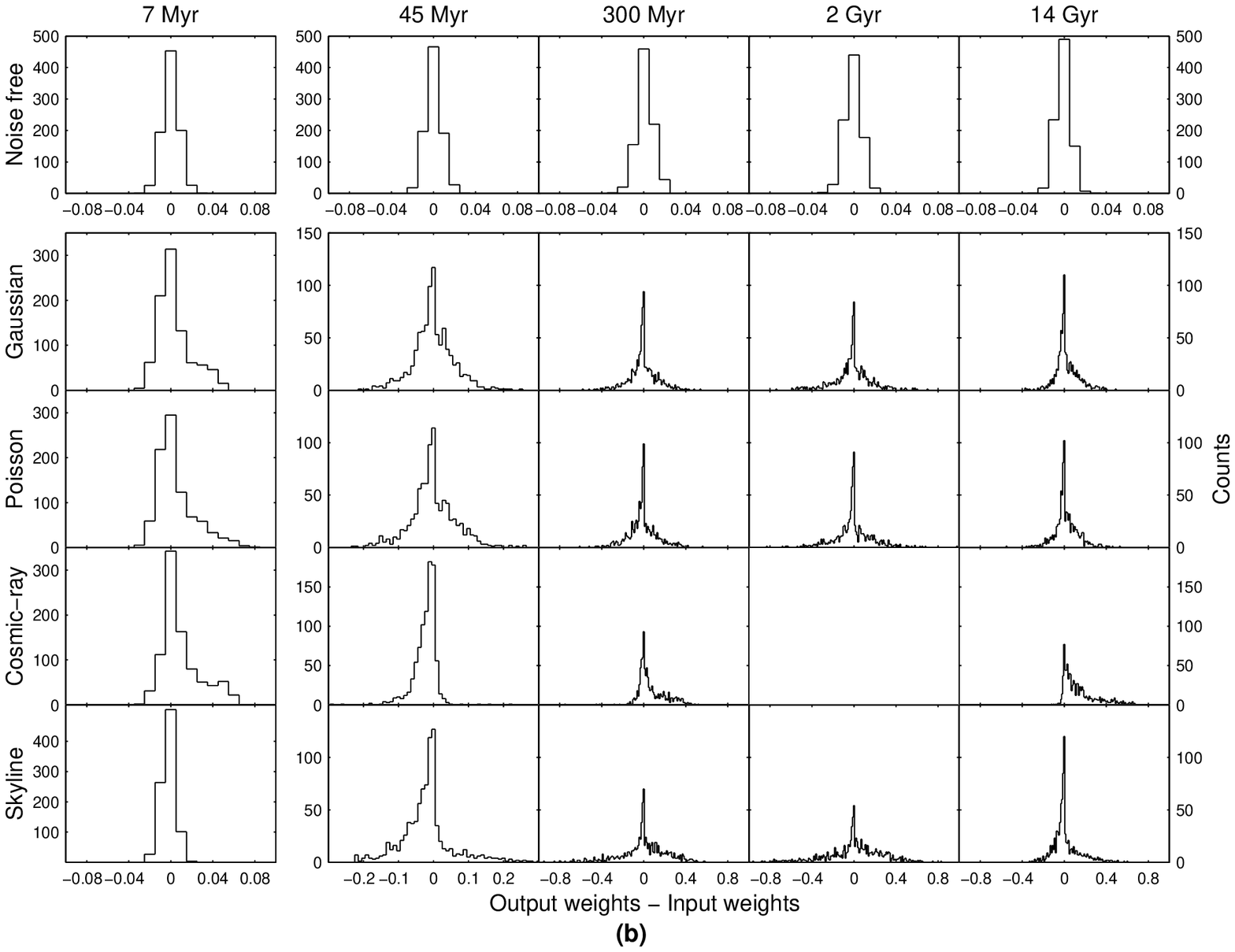,width=.5\textwidth,angle=-90}}}
\caption{Correlations (a) and distributions (b) between the weights of
the fiducial spectra in the 900 model spectra and the corresponding
weights of the data represented by 5 archetypes. In both figures the
five columns correspond to the five fiducial spectra (star formation
episodes). \textit{Row 1:} Noise free case; \textit{row 2:}
Gaussian noise added (S/N=10); \textit{row 3:} Poisson noise added
(S/N=10); \textit{row 4:} cosmic ray added (5\%); \textit{row 5:}
Skyline profile added ($\pm$5\%).}
\label{fig:ALPHAs}
\end{figure*}

Having generated sets of composite spectra we then apply archetypal
analysis. Sample sizes of 50, 100 and 900 composite spectra have been
explored. They show comparable results, and we present only the 
analysis of 900 composite spectra.  

\subsection{The number of archetypes}
We first explore the use of different numbers of archetypes. To
illustrate the key results, Fig. 2 shows a
cloud of data points corresponding to 900 composite spectra projected
onto the $(B-V,U-B)$ two-colour plane.

The corresponding projections of the five fiducial spectra are also
shown. These are identical, to machine precision, to the projections of
the archetypes when {\it five} archetypes are used.
The projection of the archetype convex hull on the two-colour plane
can be determined by noting that each edge of this polytope is a
mixture of the archetypes lying at either end, with no contribution
from the other three archetypes. The dashed line traces this
projection of the 5-archetype convex hull. Its curvature arises from
the non-linear relationship between spectral energy distribution and 
colour.

When {\it four} archetypes are used to represent the data, two of the
archetypes remained essentially unchanged (7~Myr and 45~Myr) while the
other two are mixtures of the three remaining fiducial spectra in the
forms 81\%~300 Myr + 19\% 2~Gyr and 14\% 2~Gyr + 85\% 14~Gyr.  When
{\it three} archetypes are used, the 7~Myr fiducial spectrum is
recovered, while the other two are the intermediate representations
80\% 45~Myr + 20\% 300~Myr and 44\% 2~Gyr + 56\% 14~Gyr. The
four archetype and three archetype representations produce the
archetype convex hulls shown as projections in
Fig. 2.

In summary, the use of 5 archetypes recovers exactly the 5 fiducial
spectra. When fewer archetypes are used, one or more of them represents
the ``younger'' fiducial spectra with high purity, while the other
archetypes appear as intermediate representations of the older fiducial
spectra. This result is due primarily to the presence of prominent
spectral features (especially emission lines) in the younger spectra,
and the absence of marked differences between the older fiducial
spectra.

\subsection{Data representation in the presence of noise}
Archetypal analysis is designed to focus attention on the outliers
of the data set \cite{Cutler94}. This emphasis on outliers raises the 
question of sensitivity to noise, which potentially makes a larger 
contribution in the outliers of the data set. We examine this question 
in this section, showing that archetypal analysis is robust in the 
presence of the types of noise characterising astronomical spectroscopy.

We explore the effects of noise by adding four distinct kinds of
contamination to the sample of 900 synthetic spectra described
above. They are (i) random noise with Gaussian statistics, (ii) random
noise with Poisson statistics, (iii) narrow, strong contamination
dubbed ``cosmic rays'' and (iv) a pervasive fixed pattern dubbed ``sky
subtraction residual''. 

The amplitude of the random noise models was selected to provide data 
with a characteristic signal-to-noise (S/N) ratio of 10. In particular,
for Poisson random noise, the mean noise amplitude added to a any 
wavelength bin is determined from the flux of that wavelength bin. 
Here we have ignored sky spectrum contributions.

We simulated a 5\% cosmic ray contamination rate, assuming Poisson 
statistics for the number of contaminating events per synthetic spectra. 
In this way, 44/900 spectra suffered
single-pixel events, represented as a spike of three times the 
amplitude of the peak flux density in the corresponding spectrum. In one 
spectrum (1/900) two spikes were added. The wavelength of each spike 
was chosen at random.

Sky subtraction residuals were simulated using the sky spectral
profile published by \scite{K92}. In the analysis of real spectra, sky
lines would be quasi-randomly redshifted relative to the galaxy
lines, an effect we model by randomly blueshifting the sky-line
spectrum with the range $0 \leq z \leq 0.1$. The shifted sky spectra were
rebinned in wavelength and scaled to represent random sky subtraction
errors with relative amplitudes of $\pm 5\%$ (uniform distribution), 
and added to the synthetic spectra. The combination of random redshifts 
and wavelength rebinning reduces the wavelength span from 174 to 139 
spectral bins.

Fig. 3 and Fig. 4 illustrate
respectively the effects of noise on (a) the archetypes themselves,
and (b) the coefficients that represent the data in terms of the
archetypes. In Fig. 3, the first five panels of the top
row display the fiducial spectra, which are essentially identical to
the five archetypes found in the absence of noise. Beneath this row, 
each column shows the corresponding archetype derived from the 
contaminated data set, as well as a difference spectrum with respect to 
fiducial spectrum.

With the exception of the ``2 Gyr, cosmic rays'' case which is discussed 
below, there is an evident correlation between the ``input'' and 
``derived'' weights, but there is also substantial scatter. 
Some of this scatter reflects the propagation of error into the 
archetypes themselves, although the recovered archetypes are generally 
quite similar to those derived in the absence of contamination. Most of 
the scatter arise from the propagation of error in fitting each spectrum 
as a mixture of archetypes. The appearance of large scatter for the 
``older'' spectral components reflects the fact that the three 
``older'' archetypes are similar to one another.

The rightmost panel in the top row of Fig. 3 illustrates
one of the 900 synthetic spectra. In the column below, the
corresponding contaminated synthetic spectra are exhibited, as well as
the difference from the uncontaminated spectrum. The chosen spectrum
is one that suffers a single cosmic ray event, in that form of
contamination. To the right of these panels are the corresponding
spectra obtained as the weighted sum of the archetypes. In all cases,
the archetypal representation preserves spectral detail while reducing
the amplitude of the contamination in the original spectrum. This
result confirms the potential efficacy of archetypal analysis for
compacting the size of a data set.

The results of archetypal analysis on the inferred statistical
properties of the entire sample of 900 spectra is illustrated in
Fig. 4. In part 4(a), each small panel is a scatter
plot of points whose abscissa are the weights applied to the fiducial
spectra in forming synthetic spectra, and whose ordinates are the
derived weights of the corresponding archetypal representation.
Deviations from perfect correlation reflect the propagation of errors
due to contamination of the spectra into the inferred star formation
history of each model galaxy. In Fig. 4(b), the 
corresponding distribution functions of the difference between the 
``input'' and ``derived'' weight are exhibited.

The results illustrated in Fig. 4 reveal both the
strengths and weaknesses of archetypal analysis. The tight
correlations revealed in the top row reflect the essentially perfect
correspondence between the fiducial spectra and the archetypes when
uncontaminated data are analysed using five archetypes. The
relatively tight correlations in the first two columns show that even
in the presence of significant contamination, the spectral trace of
``younger'' populations can be measured with high precision. 
The distribution of points in the 300~Myr to 14~Gyr columns reflects 
the difficulty of extracting accurately the spectral trace of single, 
old star formation events in data having relatively poor S/N ratio. 
This problem is well understood, and the results are unlikely to be 
inferior to those that might derived from other methods.

It is illuminating to interpret these results in terms of the topology
of the data and archetype convex hulls. The data can be regarded as
900 points (the number of spectra) in 174- (or 139-) dimensional (the 
number of wavelengths) space. When weak or moderate levels of Gaussian 
or Poisson noise are present, or a weak sky spectrum suffering random
wavelength shifts are present, the data points including the vertices 
of the data convex hull are displaced by distances that are relatively 
small compared with their median distance from the origin. Consequently, 
the archetype convex hull differs only mildly from the corresponding
archetype convex hull in the absence of contamination. The archetypes
resemble those of the uncontaminated data, and because the number of
archetypes is far smaller than the number of wavelengths, data
representations are significantly smoothed. 

The ``2 Gyr, cosmic rays'' panels in Fig. 3 and Fig. 4 are 
different from all other panels. In this case, archetypal 
analysis has derived as one of the archetypes a spectrum with two 
cosmic ray events in it. Because no other spectrum in the sample 
contains this pattern, the archetype does not contribute significantly to
any of the data representations.  The analysis thus yields results that closely 
approximate those that would be found were \textit{four} archetypes 
used. This case illustrates the sensitivity of archetypal analysis 
to outliers with strong and unusual noise characteristics. However, it 
also shows that the method is robust even in the presence of such noise, 
and of course reveals a way to identify and excise noisy data in a systematic
manner.

\section{Prospects}
In view of its emphasis on representing data sets as mixtures of pure
types, archetypal analysis offers considerable promise in analysing
galaxy spectra that are, presumably, superpositions of the spectra
from different emission processes (AGN, nebulae, stars) and stellar
populations of different ages. It provides near-perfect accuracy and
precision in extracting star formation histories from noise-free
synthetic data. The presence of noise, particularly with strongly
non-Gaussian statistics, compromises this high accuracy but our
studies suggest that the technique can be applied with success to
galaxy spectral data with typical noise properties.

There are extensions to archetypal analysis that we have explored in a 
limited way. For example, the results of archetypal analysis
depend to some extent on the pre-conditioning or standardisation of
the input data. We found that with cosmic ray contamination, the
best results are obtained when the data are standardised, while this
seems unnecessary with Gaussian or Poisson random noise and for the
sky subtraction residual set. We have shown that cosmic ray events may
be effectively identified through their obvious presence in some
archetypes. This invites exploration of an iterative approach to data
conditioning in which non-physical components of the archetypes are
progressively edited from the data set, providing a potentially
powerful new methodology for peeling away contaminating components in
the data. Archetypal analysis was designed to seek a representation 
using members of the data set. However, in some cases there are 
preferred \textit{models} for the data. We have confirmed that Archetype 
Analysis is still applicable in this case, with the models replacing 
the archetypes. Used in this way, Archetypal Analysis shares many of the 
features of MOPED.

We are currently using archetypal analysis to analyse real galaxy
spectra from the \scite{K92} sample and the 2dFGRS survey
\cite{Colless01}. 


\section*{Acknowledgements}
We thank Matthew Colless, Nicole Bordes and Bernard Pailthorpe for
advice and comments.



\end{document}